# Self-oscillatory interactions of supersonic streams with cylinders, placed in open channels

V. I. Pinchukov

Siberian division of Russian Academy of Sc., In-te of Computational Technologies, Novosibirsk, Russia

E-mail-address: pinchvi@ict.nsc.ru

**Abstract:** Interactions of supersonic uniform streams with cylindrical bodies, placed in open channels, are studied. Channels of rotation with the interval of cross-sectional area decreasing are considered. Two-dimensional Euler equations are solved by an implicit third order Runge-Kutta scheme. Self-oscillatory regimes are found in CFD studies at stream Mach numbers of 3 to 4.5.

## 1. Introduction

Resent paper is devoted to a search for new self-oscillatory flows. This search is carried out on a base of the hypothesis [1], that self-oscillations appear as a result of resonance interactions of "active" elements of flows, namely, elements, which amplify disturbances. Supposition is used about two types of "active" elements – contact discontinuities and intersection points of shocks with shocks or shocks with contact discontinuities. Possibility of the disturbances amplification by contact discontinuities is a result of the Kelvin-Gelmgoltce instability and is accepted. Inclusion of intersection points to a list of amplifiers is proposed in [1] as a hypothesis, which is checked by results of a search for new unsteady flows.

Self-oscillatory compressible flows may be classified into some families: 1. Flows near supersonic jets, inflowing to forward facing cavities (see, for example, [2-4]); 2. Jet impinging on a plate [1,5-9]; 3. Flows past forward-facing cavities [10-12] (outer stream is uniform in this case); 4. Cavity flows (tangential flows near surfaces with cavities) [13-16]; 5. Flows past snaked bodies [1,17-19]; 6. Flows past bluff bodies (vortex shedding from bluff bodies may result unsteady wake structures) [20-22]; 7. Transonic flows near profiles (Euler equations may have non-unique solutions in this case, which results bifurcations and self –oscillations) [23-24].

Numerical investigations of flows, containing the most number of "active" elements, are used to search for new unsteady flows in [25-27]. Flows near blunted bodies (cylinders or cones), giving off opposite jets, are discovered to have intensive self-sustained oscillations [25-26]. These flows may not be included to any mentioned above class of unsteady flows. Sonic underexpanded jet impinging on the pair open tube – inner cylinder [27] are found to have self-oscillatory regimes. It seems that these flows are closed to second class (see above) of unsteady flows.

A search for self-oscillatory flows is continued here. Supersonic flows near the pair open channel – inner cylinder are considered. These flows may contain shock waves, contact discontinuities, intersection points. CFD studies of these flows are carried out and self-oscillations are observed for stream Mach numbers $3 \leq M_\infty \leq 4.5$.

## 2. CFD Design approach

**2.1. Boundary conditions.** Fig. 1 represents schematically a numerical domain and a mesh near a cylindrical body, placed in an open channel. All variables are prescribed at the inflow boundary (*HA*). Parameters of the uniform stream are set at this boundary, namely, Mach number $M=M_\infty$, density $\rho=1$, pressure $p=1$ (in dimensionless form), the radial velocity $v=0$. The normal velocity is equal to zero and other variables are extrapolated at solid surfaces (*CB,CD,FE,FG*). The radial velocity $v=0$ at the symmetry axis *HG*, other variables are extrapolated. Extrapolation conditions are used at the tube exit *DE* and at the *AB* boundary.

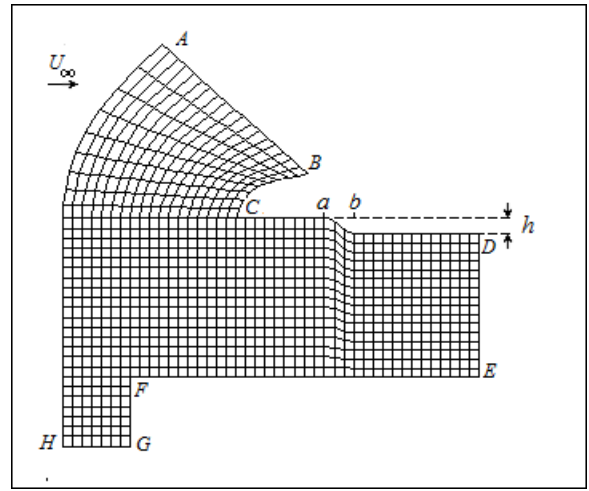

**Fig. 1**. Schematic representation of a numerical domain and a mesh.

The channel form at the [a,b] interval of cross-sectional area decreasing is defined by the formulae

$$Y(x) = R_{tub} - 16h(x-a)^2 (x-2b+a)^2 / (b-a)^4 .$$

**2.2. Numerical method.** An implicit conservative Runge-Kutta scheme [28] is modified and employed here. Namely, a special version of the code is developed for the case when functions $x=x(a,b)$, $y=y(a,b)$ perform mapping of the unit square with excisions $\{0 \leq a \leq a_0, 0 \leq b \leq b_0\}$, $\{a_1 \leq a \leq 1, 0 \leq b \leq b_1\}$ to a curvilinear quadrangle with curvilinear quadrangular excisions (see fig. 1). This version allows carrying out calculations, described below, without dividing complicated domains into subdomains. Both recent method and method [28] are third order (viscous terms are approximated with second order). The $515 \times 586$ mesh is used in written below calculations.

Naturally, numerical calculations deal with dimensionless variables. These variables are defined as relations of initial variables and next parameters of the outer stream or the body size: $p_\infty$ - for pressure, $\rho_\infty$ - for a density, $\sqrt{p_\infty/\rho_\infty}$ - for a velocity, $r_{tub}$=y(C)-y(H) (the maximum inner channel radius) – for space variables, $r_{tub}/\sqrt{p_\infty/\rho_\infty}$ - for time.

## 3. Results and discussion

A search for self-oscillatory flows is carried out for each stream Mach number, considered below, by trial calculations of some variants. It is observed, that self-oscillations may appear, if the relation of channel and cylinder lengths provides position of shock waves intersection points closed to the channel edges. All flows considered below contain separation zones, starting at cylinder edges (signed by *F* in fig.1). A search is stopping, if 1-3 unsteady flows are found. Examples of unsteady flows for different Mach numbers are represented below.

**3.1. Stream Mach number $M_\infty$=3.** Figs. 2a and 2b shows density distributions at different instants in the flow, calculated for the geometry parameters $L_{cyl}$=1.4 (the cylinder length), $R_{cyl}$=.3 (the cylinder radius), $L_{tub}$=0.9 (the channel length), $R_{min}$ = $R_{tub}$ -h=1-h=.91 (the least channel radius), $h$ =0.09, $a$=x(F)+ 0.25 $L_{tub}$, $b$=a+0.1 (see fig. 1). Solid walls (cylinder and channel walls) are shown by bold lines.

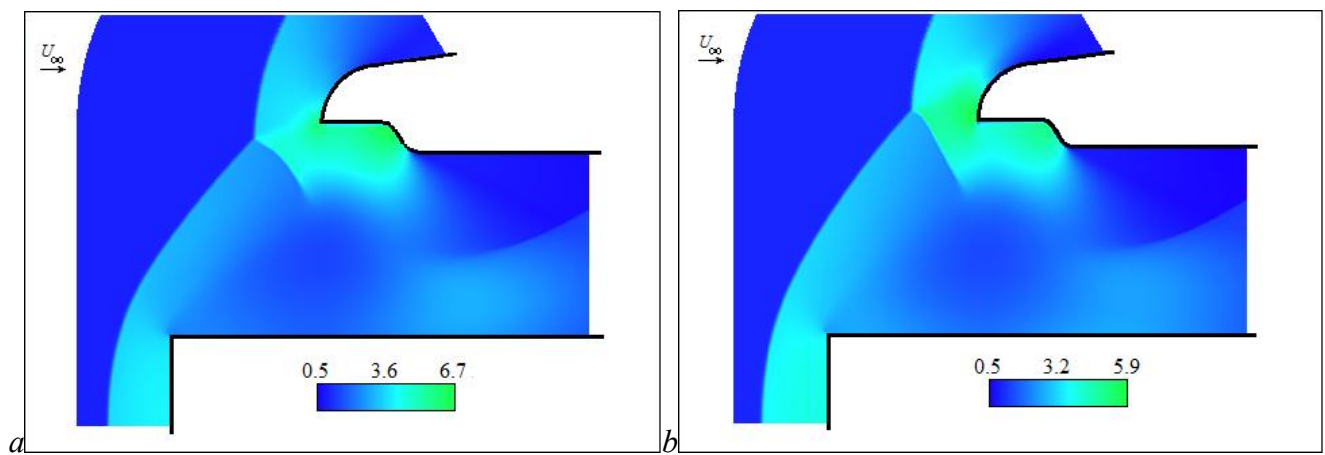

**Fig. 2**. Density distributions, $M_\infty$=3, a - t=34.1, b - t=34.1+T/2.

Density magnitudes at the cylinder edge (point C in fig. 1), are plotted in fig. 3.

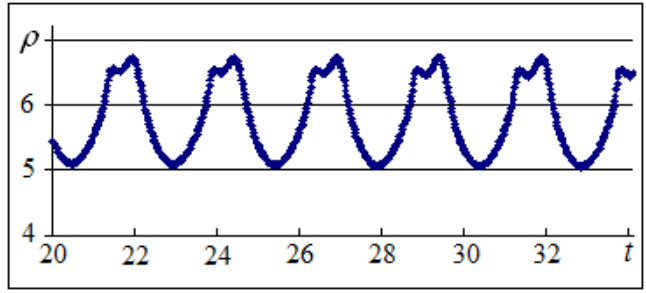

**Fig. 3.** The density history, $M_\infty=3$.

The density history, represented in fig. 3, shows that this flow is nearly periodic with the $T=2.47$ period. Fig. 2a corresponds to the final time instant $t=34.1$ (see fig. 3). To illustrate the flow dynamic through one period $T$ the density distribution for the instant $t=34.1+T/2$ is shown in fig. 2b. If to compare figs. 2a and 2b, different position of the shock waves intersection point may be seen. The most intensive flow oscillations are observed in the region between this intersection point and the channel edge (signed by $C$ in fig.1).

**3.2. Stream Mach number $M_\infty=3.5$.** Fig. 4 shows the density history for the self-oscillatory flow, defined by geometry parameters $L_{cyl}=1.4$ (the cylinder length), $R_{cyl}=.3$ (the cylinder radius), $L_{tub}=0.9$ (the channel length), $R_{min} = R_{tub}-h=1-h=.92$ (the least channel radius), $h=.08$, $a=x(C)+ 0.25L_{tub}$, $b=a+0.1$ (see fig. 1). Density magnitudes at the point C are plotted in this fig.

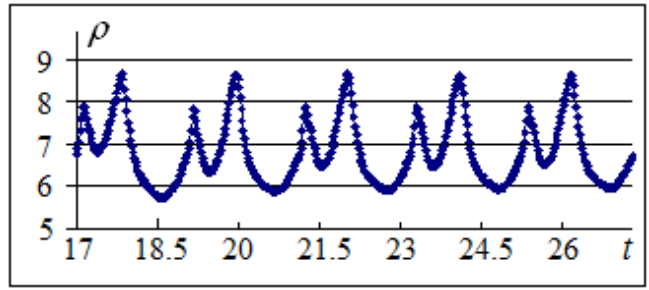

**Fig. 4**. The density history, $M_\infty=3.5$.

The density history, presented in fig. 4, illustrates that this flow is nearly periodic with the $T=2.19$ period. If to compare periods for Mach numbers 3.0 and 3.5, decreasing of the self-oscillation period is seen when Mach number is increasing. This tendency is true for next considered flows. To show the flow dynamic through one period $T$ density distributions for instants $t=27.4+T/8$ and $t=27.4+T5/8$ are represented in figs. 5a and 5b.

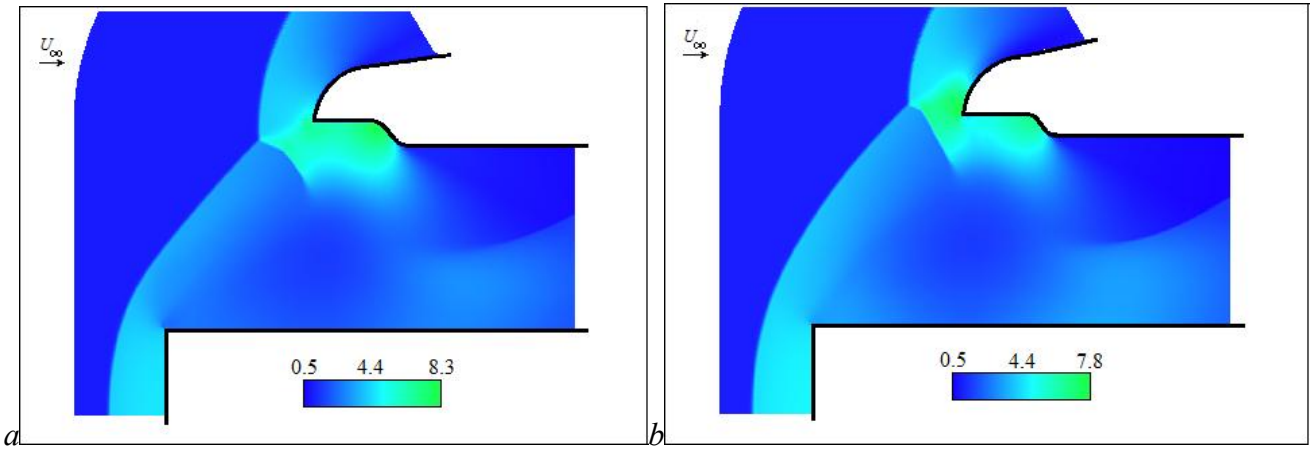

**Fig. 5.** Density Distributions, *a* - *t*=27.4+*T*/8, *b* - *t*=27.4+5*T*/8.

If to compare figs. 5*a* and 5*b*, different position of the shock waves intersection point may be seen.

**3.3. Stream Mach number $M_\infty$=4.** Figs. 6*a* and 6*b* show density distributions for two instants. The flow is calculated for geometry parameters $L_{cyl}$=1.4, $R_{cyl}$=.3, $L_{tub}$=0.8, $R_{min}$ = $R_{tub}$ -h= 1-h=.96, h =.04, a=x(C)+ 0.25$L_{tub}$, b=a+0.1 (see fig. 1).

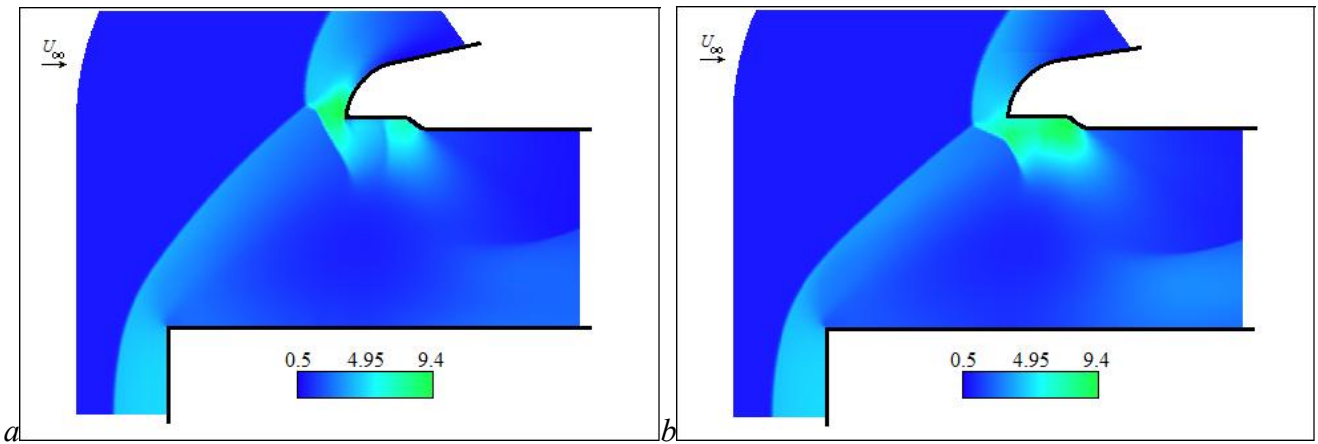

**Fig. 6.** Density distributions, *a* - *t*=31.1, *b* - *t*=31.1+*T*/2

Density magnitudes at the cylinder edge (point C in fig. 1) are plotted in fig. 7.

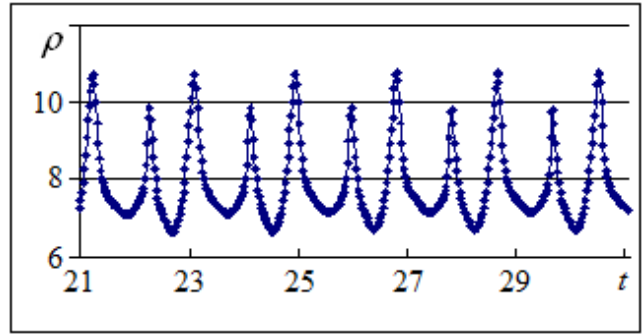

**Fig. 7**. The density history, $M_\infty = 4$.

So, this flow is nearly periodic with the $T=1.88$ period. Flow fields dynamics during one period after the final instant t=31.1 (see fig. 7) is calculated. Density distributions are shown for instants $t=31.1$ (fig. 6a) and $t=31.1+T/2$ (fig. 6b), correspondingly. Comparison of figs. 6a and 6b allows to conclude, that these figs. differ one from another by position of the shock waves intersection point.

**3.4. Stream Mach number $M_\infty=4.5$.** Fig. 8 shows the history of density magnitudes at point $C$ for the self-oscillatory flow at $M_\infty = 4.5$. Calculations are carried out for geometry parameters $L_{cyl}=1.4$, $R_{cyl}=.3$, $L_{tub}=0.8$, $R_{min} = R_{tub} - h = 1 - h = .94$, $h=.06$, $a=x(C)+0.25L_{tub}$, $b=a+0.1$ (see fig. 1).

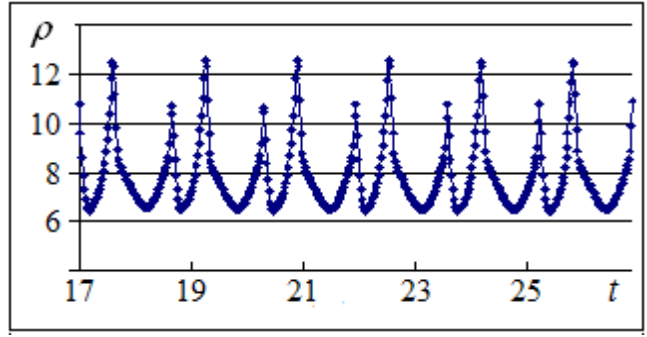

**Fig. 8.** The density history, $M_\infty = 4.5$.

According to the density history, presented in fig. 8, this flow is nearly periodic with the $T=1.65$ period. Flow fields dynamics during one period after the final instant $t=27.6$ (see fig. 8) is calculated. Density distributions are shown in fig. 9 for time instants $t=27.6+T/8$ (fig. 9a) and $t=27.6+T5/8$ (fig. 9b), correspondingly. It can be seen again that these figs. differ one from another by position of the shock waves intersection point.

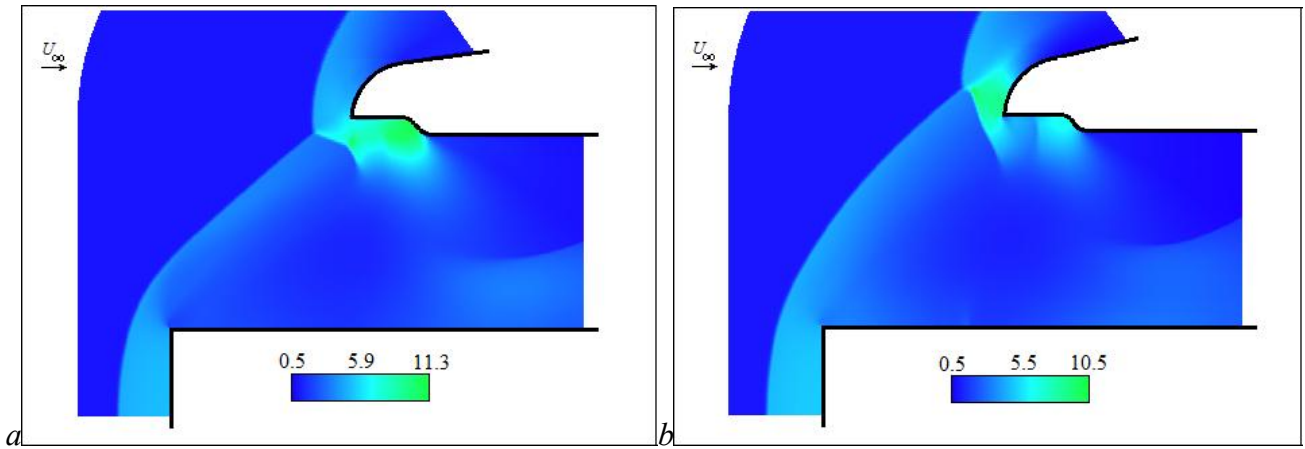

**Fig. 9.** Density distributions, *a* - *t*=27.6+*T*/8. *b* - *t*=27.6+*T*5/8

The most intensive flow oscillations are observed in the region between the intersection point and the channel edge (signed by *C* in fig.1).

## 4. Conclusions

Recent paper is devoted to CFD search for new self-oscillatory compressible flows. Flows with the most number of "active" elements are investigated numerically. Interactions of supersonic uniform streams with cylindrical bodies, placed in open channels, are studied. These flows contain several "active" elements, so unsteady regimes are possible according to the written above hypothetical mechanism of self-oscillations. Unsteady flows are observed at stream Mach numbers $3 \leq M_\infty \leq 4.5$. The relation of inner cylinder and channel lengths should provide position of the shock waves intersection point closed to channel edge (signed by *C* in fig. 1). Channels with the interval of cross-sectional area decreasing are considered. This decreasing is conducive to possibilities of the self-oscillatory regimes appearance. If the interval of decreasing is absent, self-sustained oscillations are not observed in recent investigations.

At first glance these flows may be included to a class of unsteady supersonic flows past forward-facing cavities (the class 3, see introduction). But recent flow physics is seemed to be more similar to the flow physics of the 5th type, namely, to flows physics near snaked bodies. Last flows contain shock waves, produced by snakes. These waves move to main blunted bodies, where intensive oscillations are observed. Similarly, recent investigations show, that inner cylinders produce shock waves, which move to channel buff-ends, where the most intensive oscillations take place.

---

## Nomeclatures

*P* - pressure,

$\rho$ - density,

$M_\infty$ - stream Mach number,

$L_{cyl}$ - cylinder length,

$L_{tub}$ - tube length,

$r_{tub}$ - tube radius at initial cross section,

$r_{cyl}$ - cylinder radius,

$L_{tub}$ - tube length,

---